\documentclass[12pt,a4paper]{article}
\usepackage{amsmath}
\usepackage{graphicx}
\begin{document}
\begin{center}
{\bfseries Measurement of neutrino flux from the primary proton\textendash{}proton
fusion process in the Sun with Borexino detector. 
\footnote{\small Talk at the International Workshop on Prospects of Particle Physics: "Neutrino Physics and Astrophysics", JINR, INR, 1 February - 8 February 2015, Valday, Russia.}}
\vskip 5mm 
O.Yu.Smirnov$^{\mbox{a}}$ on behalf of the Borexino collaboration:
\vskip 5mm
M.~Agostini$^{\mbox{b}}$,
S.~Appel$^{\mbox{b}}$,
G.~Bellini$^{\mbox{c}}$,
J.~Benziger$^{\mbox{d}}$,
D.~Bick$^{\mbox{e}}$,
G.~Bonfini$^{\mbox{f}}$,
D.~Bravo$^{\mbox{g}}$,
B.~Caccianiga$^{\mbox{c}}$,
F.~Calaprice$^{\mbox{h}}$,
A.~Caminata$^{\mbox{i}}$,
P.~Cavalcante$^{\mbox{f}}$,
A.~Chepurnov$^{\mbox{j}}$,
K.~Choi$^{\mbox{k}}$,
D.~D'Angelo$^{\mbox{c}}$,
S.~Davini$^{\mbox{l}}$,
A.~Derbin$^{\mbox{m}}$,
L.~Di Noto$^{\mbox{i}}$,
I.~Drachnev$^{\mbox{l}}$,
A.~Empl$^{\mbox{n}}$,
A.~Etenko$^{\mbox{o}}$,
K.~Fomenko$^{\mbox{a}}$,
D.~Franco$^{\mbox{p}}$,
F.~Gabriele$^{\mbox{f}}$,
C.~Galbiati$^{\mbox{h}}$,
C.~Ghiano$^{\mbox{i}}$,
M.~Giammarchi$^{\mbox{c}}$,
M.~Goeger-Neff$^{\mbox{b}}$,
A.~Goretti$^{\mbox{h}}$,
M.~Gromov$^{\mbox{j}}$,
C.~Hagner$^{\mbox{e}}$,
E.~Hungerford$^{\mbox{n}}$,
Aldo~Ianni$^{\mbox{f}}$,
Andrea~Ianni$^{\mbox{h}}$,
K.~Jedrzejczak$^{\mbox{r}}$,
M.~Kaiser$^{\mbox{e}}$,
V.~Kobychev$^{\mbox{s}}$,
D.~Korablev$^{\mbox{a}}$,
G.~Korga$^{\mbox{f}}$,
D.~Kryn$^{\mbox{p}}$,
M.~Laubenstein$^{\mbox{f}}$,
B.~Lehnert$^{\mbox{t}}$,
E.~Litvinovich$^{\mbox{o}}$$^{\mbox{u}}$,
F.~Lombardi$^{\mbox{f}}$,
P.~Lombardi$^{\mbox{c}}$,
L.~Ludhova$^{\mbox{c}}$,
G.~Lukyanchenko$^{\mbox{o}}$$^{\mbox{u}}$,
I.~Machulin$^{\mbox{o}}$$^{\mbox{u}}$,
S.~Manecki$^{\mbox{g}}$,
W.~Maneschg$^{\mbox{v}}$,
S.~Marcocci$^{\mbox{l}}$,
E.~Meroni$^{\mbox{c}}$,
M.~Meyer$^{\mbox{e}}$,
L.~Miramonti$^{\mbox{c}}$,
M.~Misiaszek$^{\mbox{r}}$$^{\mbox{f}}$,
P.~Mosteiro$^{\mbox{h}}$,
V.~Muratova$^{\mbox{m}}$,
B.~Neumair$^{\mbox{b}}$,
L.~Oberauer$^{\mbox{b}}$,
M.~Obolensky$^{\mbox{p}}$,
F.~Ortica$^{\mbox{w}}$,
K.~Otis$^{\mbox{x}}$,
L.~Pagani$^{\mbox{i}}$,
M.~Pallavicini$^{\mbox{i}}$,
L.~Papp$^{\mbox{b}}$,
L.~Perasso$^{\mbox{i}}$,
A.~Pocar$^{\mbox{x}}$,
G.~Ranucci$^{\mbox{c}}$,
A.~Razeto$^{\mbox{f}}$,
A.~Re$^{\mbox{c}}$,
A.~Romani$^{\mbox{w}}$,
R.~Roncin$^{\mbox{f}}$$^{\mbox{p}}$,
N.~Rossi$^{\mbox{f}}$,
S.~Sch\"onert$^{\mbox{b}}$,
D.~Semenov$^{\mbox{m}}$,
H.~Simgen$^{\mbox{v}}$,
M.~Skorokhvatov$^{\mbox{o}}$$^{\mbox{u}}$,
A.~Sotnikov$^{\mbox{a}}$,
S.~Sukhotin$^{\mbox{o}}$,
Y.~Suvorov$^{\mbox{y}}$$^{\mbox{o}}$,
R.~Tartaglia$^{\mbox{f}}$,
G.~Testera$^{\mbox{i}}$,
J.~Thurn$^{\mbox{t}}$,
M.~Toropova$^{\mbox{o}}$,
E.~Unzhakov$^{\mbox{m}}$,
R.B.~Vogelaar$^{\mbox{g}}$,
F.~von~Feilitzsch$^{\mbox{b}}$,
H.~Wang$^{\mbox{y}}$,
S.~Weinz$^{\mbox{z}}$,
J.~Winter$^{\mbox{z}}$,
M.~Wojcik$^{\mbox{r}}$,
M.~Wurm$^{\mbox{z}}$,
Z.~Yokley$^{\mbox{g}}$,
O.~Zaimidoroga$^{\mbox{a}}$,
S.~Zavatarelli$^{\mbox{i}}$,
K.~Zuber$^{\mbox{t}}$,
G.~Zuzel$^{\mbox{r}}$.

\vskip 5mm
{\small {\it $^{\mbox{a}}$ Joint Institute for Nuclear Research, 141980 Dubna, Russia.}}\\
{\small {\it $^{\mbox{b}}$ Physik-Department and Excellence Cluster Universe, Technische Universit\"at  M\"unchen, 85748 Garching, Germany.}}\\
{\small {\it $^{\mbox{c}}$ Dipartimento di Fisica, Universit\`a degli Studi e INFN, 20133 Milano, Italy.}}\\
{\small {\it $^{\mbox{d}}$ Chemical Engineering Department, Princeton University, Princeton, NJ 08544, USA.}}\\
{\small {\it $^{\mbox{e}}$ Institut f\"ur Experimentalphysik, Universit\"at, 22761 Hamburg, Germany.}}\\
{\small {\it $^{\mbox{f}}$ INFN Laboratori Nazionali del Gran Sasso, 67010 Assergi (AQ), Italy.}}\\
{\small {\it $^{\mbox{g}}$ Physics Department, Virginia Polytechnic Institute and State University, Blacksburg, VA 24061, USA.}}\\
{\small {\it $^{\mbox{h}}$ Physics Department, Princeton University, Princeton, NJ 08544, USA.}}\\
{\small {\it $^{\mbox{i}}$ Dipartimento di Fisica, Universit\`a degli Studi e INFN, Genova 16146, Italy.}}\\
{\small {\it $^{\mbox{j}}$ Lomonosov Moscow State University Skobeltsyn Institute of Nuclear Physics, 119234 Moscow, Russia.}}\\
{\small {\it $^{\mbox{k}}$ Department of Physics and Astronomy, University of Hawaii, Honolulu, HI 96822, USA.}}\\
{\small {\it $^{\mbox{l}}$  Gran Sasso Science Institute (INFN), 67100 \L'Aquila, Italy.}}\\
{\small {\it $^{\mbox{m}}$ St. Petersburg Nuclear Physics Institute NRC Kurchatov Institute, 188350 Gatchina, Russia.}}\\
{\small {\it $^{\mbox{n}}$ Department of Physics, University of Houston, Houston, TX 77204, USA.}}\\
{\small {\it $^{\mbox{o}}$ NRC Kurchatov Institute, 123182 Moscow, Russia.}}\\
{\small {\it $^{\mbox{p}}$ AstroParticule et Cosmologie, Universit\'e Paris Diderot, CNRS/IN2P3, CEA/IRFU, Observatoire de Paris, Sorbonne Paris Cit\'e, 75205 Paris Cedex 13, France.}}\\
{\small {\it $^{\mbox{r}}$ M.~Smoluchowski Institute of Physics, Jagiellonian University, 30059 Krakow, Poland.}}\\
{\small {\it $^{\mbox{s}}$ Kiev Institute for Nuclear Research, 06380 Kiev, Ukraine.}}\\
{\small {\it $^{\mbox{t}}$ Department of Physics, Technische Universit\"at Dresden, 01062 Dresden, Germany.}}\\
{\small {\it $^{\mbox{u}}$  National Research Nuclear University MEPhI (Moscow Engineering Physics Institute), 115409 Moscow, Russia.}}\\
{\small {\it $^{\mbox{v}}$ Max-Planck-Institut f\"ur Kernphysik, 69117 Heidelberg, Germany.}}\\
{\small {\it $^{\mbox{w}}$ Dipartimento di Chimica, Biologia e Biotecnologie, Universit\`a e INFN, 06123 Perugia, Italy.}}\\
{\small {\it $^{\mbox{x}}$ Amherst Center for Fundamental Interactions and Physics Department, University of Massachusetts, Amherst, MA 01003, USA.}}\\
{\small {\it $^{\mbox{y}}$ Physics and Astronomy Department, University of California Los Angeles (UCLA), Los Angeles, California 90095, USA.}}\\
{\small {\it $^{\mbox{z}}$ Institute of Physics and Excellence Cluster PRISMA, Johannes Gutenberg-Universit\"at Mainz, 55099 Mainz, Germany.}}\\
\end{center}
\vskip 5mm
\centerline{\bf Abstract}
Neutrino produced in a chain of nuclear reactions in the Sun starting from the 
fusion of two protons, for the first time has been detected in a 
real-time detector in spectrometric mode. The unique properties of the 
Borexino detector provided an oppurtunity to disentangle pp-neutrino spectrum 
from the background components. A comparison of the total neutrino flux from 
the Sun with Solar luminosity in photons provides a test of the stability of 
the Sun on the 10$^{5}$ years time scale, and sets a strong limit on the
power production in the unknown energy sources in the Sun of no more than 4\% of 
the total energy production at 90\% C.L.
\section{Introduction}

The solar photon luminosity (a total power radiated in form of photons
into space) is determined by measuring the total solar irradiance
by spacecrafts over the wide subrange of the electromagnetic spectrum,
from x-rays to radio wavelengths; it has been accurately monitored
for decades. The luminosity $L_{\odot}=3.846\times10^{26}$ W is measured
for a precision of 0.4\% with the largest uncertainty of about 0.3\%
due to disagreements between the measurements of different satellite
detectors~\cite{Chapman,SunTotal}. The energy lost by neutrinos
adds $L_{\nu}=0.023\cdot L_{\odot}$ to this value~\cite{Bahcall1989}.
The solar luminosity constraint on the solar neutrino fluxes can be
written as~\cite{LumConstr}:

\begin{equation}
\frac{L_{\odot}}{4\pi(1a.u.)^{2}}=\sum\alpha_{i}\Phi_{i}\label{Lum}
\end{equation}

where 1 a.u. is the average earth-sun distance, the coefficient $\alpha_{i}$
is the amount of energy provided to the star by nuclear fusion reactions
associated with each of the important solar neutrino fluxes, $\Phi_{i}$.
The numerical values of the $\alpha$\textquoteright{}s are determined
to an accuracy of $10^{-4}$ and better. 

The estimated uncertainty in the luminosity of the Sun corresponds
to less than 3\% uncertainty in total solar neutrino flux.

The Sun is a weakly variable star, its luminosity has short term fluctuations~\cite{SunTotal,SolarVars}. The major fluctuation occurs during the
eleven-year solar cycle with amplitude of about 0.1\%. Long-term solar
variability (such as the Maunder minimum in the 16th and 17th century)
is commonly beleived to do not exceed the short term variations. 

Because of the relation (\ref{Lum}) between the solar photon and
neutrino luminosity, the measurement of the total neutrino luminosity
will provide a test of the stability of the Sun at the time scale
of 40000 years~\cite{FR99}, the time needed for the radiation born
at the center of the Sun to arrive to its surface. Finding a disagreement
between $L_{\odot}$ and $L_{\nu}$ would have significant long term
enviromental implications, and in the case of an agreement of two
measurements it would be possible to limit the unknown sources of
the solar energy, different from the known thermonuclear fusion of
light elements in the pp-chain and CNO-cycle.

The main neutrino sources in the Sun are the pp- and $^{7}$Be reactions,
providing roughly 91 and 7\% of the total neutrino flux respectively.
Borexino already measured $^{7}$Be neutrino flux with $5\%$ precision~\cite{Be11}, 
but till recent time the pp-neutrino flux was derived
in a differential measurement using the data of solar detectors. 

Solar pp neutrinos measurement is a critical test of stellar evolution
theory, discussion of the physics potential of the pp solar neutrino
flux measurement can be found in~\cite{pp_potential,WhyPP,Raghavan}
(at the moment of the discussion the parameters space for MSW solution
was not established yet, thus the authors were giving priority to
this part of the physical potential). 

A number of projects aiming to perform pp-neutrino detection have
been put forward in past two decades, but with all the time passed
since the proposals, none of them started the operation facing the
technical problems with realization. The principal characteristics
of the proposals are presented in table\ensuremath{\:}
\ref{Tab:Experiments}. The radiochemical experiments sensitive
to the solar pp- neutrinos (SAGE~\cite{SAGE09} and GALLEX~\cite{Gallex})
are not cited in the table, the combined best fit of the radiochemical
and other solar experiments gives solar pp-neutrino flux of $(6.0\pm0.8)\times10^{10}$
cm$^{-2}$s$^{-1}$~\cite{SAGE09} in a good agreement with expected
value of $6.0\times(1.000\pm0.006)\times10^{10}$ cm$^{-2}$s$^{-1}$.

\begin{table*}[!th]
\tiny{
\begin{centering}
{\scriptsize }%
\begin{tabular}{|c|c|c|c|c|l|c|}
\hline 
Project%
 & %
Method%
 & %
Threshold%
 & %
Resolution %
 & %
Mass {[}t{]}%
 & \multicolumn{1}{c|}{%
Reaction%
} & %
pp%
\tabularnewline
(reference)%
 & %
 & %
{[}keV{]}%
 & %
 & %
 & %
 & %
events%
\tabularnewline
 & %
 & %
 & %
 & %
 & %
 & %
{[}\ensuremath{d^{-1}}
{]}%
\tabularnewline
\hline 
LENS%
 & %
\ensuremath{^{176}Yb}
, %
 & %
301 (\ensuremath{\nu}
)%
 & %
7\%%
 & %
20%
 & \multicolumn{1}{l|}{%
\ensuremath{^{176}Yb+\nu_{e}\rightarrow}
} & %
0.5%
\tabularnewline
\cite{LENS}%
 & %
LS%
 & %
 & %
@ 1 MeV%
 & %
(8\% nat \ensuremath{^{176}Yb}
)%
 & \multicolumn{1}{r|}{%
\ensuremath{^{176}Lu+e^{-}}
} & %
\tabularnewline
\hline 
INDIUM%
 & %
\ensuremath{^{115}In}
 & %
118(\ensuremath{\nu}
)%
 & %
5-10\%%
 & %
4%
 & \multicolumn{1}{l|}{%
\ensuremath{^{115}In+\nu_{e}\rightarrow}
} & %
1.0%
\tabularnewline
\cite{Indium}%
 & %
LS%
 & %
 & %
@1 MeV%
 & %
 & \multicolumn{1}{r|}{%
\ensuremath{^{115}Sn^{*}(613)+e^{-}}
} & %
\tabularnewline
\hline 
GENIUS%
 & %
\ensuremath{^{76}Ge}
 & %
11(\ensuremath{e^{-}}
)%
 & %
0.3\% %
 & %
1 %
 & \multicolumn{1}{l|}{%
\ensuremath{\nu+e^{-}\rightarrow}
} & %
1.8%
\tabularnewline
\cite{Genius}%
 & %
scatt%
 & %
59(\ensuremath{\nu}
)%
 & %
@ 300 keV%
 & %
10%
 & \multicolumn{1}{r|}{%
\ensuremath{\nu+e^{-}}
} & %
18%
\tabularnewline
\hline 
HERON%
 & %
superfluid \ensuremath{^{4}He}
 & %
50(\ensuremath{e^{-}}
)%
 & %
8.3\%%
 & %
10%
 & \multicolumn{1}{l|}{%
\ensuremath{\nu+e^{-}\rightarrow}
} & %
14%
\tabularnewline
\cite{HERON,HERON02}%
 & %
rotons/phonons+uv %
 & %
141(\ensuremath{\nu}
)%
 & %
@364 keV%
 & %
 & \multicolumn{1}{r|}{%
\ensuremath{\nu+e^{-}}
} & %
\tabularnewline
\hline 
XMASS%
 & %
liquid Xe%
 & %
50(\ensuremath{e^{-}}
)%
 & %
17.5\%%
 & %
10%
 & \multicolumn{1}{l|}{%
\ensuremath{\nu+e^{-}\rightarrow}
} & %
14%
\tabularnewline
\cite{XMASS}%
 & %
scintill%
 & %
141(\ensuremath{\nu}
)%
 & %
@ 100 keV%
 & %
 & \multicolumn{1}{r|}{%
\ensuremath{\nu+e^{-}}
} & %
\tabularnewline
\hline 
CLEAN%
 & %
liquid Ne%
 & %
20(\ensuremath{e^{-}}
)%
 & %
 & %
135%
 & \multicolumn{1}{l|}{%
\ensuremath{\nu+e^{-}\rightarrow}
} & %
7.2%
\tabularnewline
\cite{CLEAN}%
 & %
 & %
82(\ensuremath{\nu}
)%
 & %
 & %
 & %
\ensuremath{\nu+e^{-}}
 & %
\tabularnewline
\hline 
HELLAZ%
 & %
He (5 atm), %
 & %
100(\ensuremath{e^{-}}
)%
 & %
6\%%
 & %
2000 \ensuremath{m^{3}}
 & \multicolumn{1}{l|}{%
\ensuremath{\nu+e^{-}\rightarrow}
} & %
7%
\tabularnewline
\cite{HELLAZ}%
 & %
TPC%
 & %
217(\ensuremath{\nu}
)%
 & %
@800 keV%
 & %
 & \multicolumn{1}{r|}{%
\ensuremath{\nu+e^{-}}
} & %
\tabularnewline
\hline 
MOON%
 & %
drift %
 & %
168(\ensuremath{\nu}
)%
 & %
12.4\% FWHH%
 & %
3.3%
 & \multicolumn{1}{l|}{%
\ensuremath{\nu_{e}+^{100}Mo\rightarrow}
} & %
1.1%
\tabularnewline
\cite{MOON}%
 & %
chambers%
 & %
 & %
@ 1 MeV%
 & %
 & %
\ensuremath{^{100}Tc+e^{-}}
 & %
\tabularnewline
\hline 
MUNU%
 & %
TPC,\ensuremath{CF_{4}}
 & %
100(\ensuremath{e^{-}}
)%
 & %
16\% FWHH%
 & %
0.74 %
 & \multicolumn{1}{l|}{%
\ensuremath{\nu+e^{-}\rightarrow}
} & %
0.5%
\tabularnewline
\cite{MUNU}%
 & %
direction%
 & %
217(\ensuremath{\nu}
)%
 & %
@ 1 MeV%
 & %
(200 \ensuremath{m^{3}}
)%
 & \multicolumn{1}{r|}{%
\ensuremath{\nu+e^{-}}
} & %
\tabularnewline
\hline 
NEON%
 & %
He,Ne%
 & %
20(\ensuremath{e^{-}}
)%
 & %
16\% FWHH%
 & %
10%
 & \multicolumn{1}{l|}{%
\ensuremath{\nu+e^{-}\rightarrow}
} & %
18%
\tabularnewline
\cite{NEON}%
 & %
scintill%
 & %
82(\ensuremath{\nu}
)%
 & %
@ 100 keV%
 & %
 & \multicolumn{1}{r|}{%
\ensuremath{\nu+e^{-}}
} & %
\tabularnewline
\hline 
10 t LS%
 & %
LS%
 & %
170(\ensuremath{e^{-}}
)%
 & %
10.5 keV %
 & %
10%
 & \multicolumn{1}{l|}{%
\ensuremath{\nu+e^{-}\rightarrow}
} & %
1.8%
\tabularnewline
\cite{ppCTF,ppCTF-2}%
 & %
 & %
310(\ensuremath{\nu}
)%
 & %
@ 200 keV%
 & %
 & \multicolumn{1}{r|}{%
\ensuremath{\nu+e^{-}}
} & %
\tabularnewline
\hline 
Borexino%
 & %
LS%
 & %
165(\ensuremath{e^{-}}
)%
 & %
5\%%
 & %
75.5%
 & %
\ensuremath{\nu+e^{-}\rightarrow}
 & %
13.6%
\tabularnewline
\cite{Brx14}%
 & %
 & %
305(\ensuremath{\nu}
)%
 & %
@1 MeV%
 & %
(fiducial)%
 & \multicolumn{1}{r|}{%
\ensuremath{\nu+e^{-}}
} & %
\tabularnewline
\hline 
\end{tabular}
\par\end{centering}\par
\caption{%
\label{Tab:Experiments}{Key characteristics
of the solar neutrino projects sensitive to the pp-neutrino. The number
of expected neutrino is calculated for the fraction of the neutrino
spectrum above the threshold, but the region of sensitivity (limited
i.e. by signal to backgrounds ratio of 1) could be stricter.}%
}
}

\end{table*}

A possibility to use ultrapure liquid organic scintillator as a low
energy solar neutrino detector for a first time was discussed in ~\cite{ppCTF,ppCTF-2}.
The authors come to the conclusion that a liquid scintillator detector
with an active volume of 10 tons is a feasible tool to register the
solar pp-neutrino if operated at the target level of radiopurity for
Borexino and good energy resolution (5\% at 200 keV) is achieved.

\begin{figure}[!th]

\begin{centering}
\includegraphics[width=0.5\textwidth]{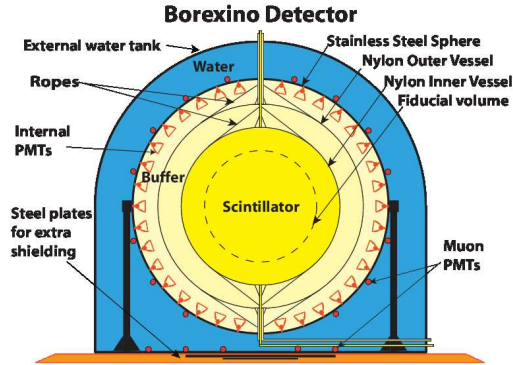}
\par\end{centering}

\centering{}\caption{\label{Borexino}Schematic view of the Borexino detector.}
\end{figure}

\section{Borexino detector}

The Borexino detector consists of a dome-like structure (see Fig.\ref{Borexino}),
16 meters in diameter, filled with a mass of 2,400 tons of highly
purified water which acts as the shield against the external radioactive
emissions of the rocks and the environment that surround the facility.
The water buffer acts also as an effective detector of the residual
cosmic rays. Within the volume of water a steel sphere is mounted
which hosts 2,200 looking inward photomultiplier tubes providing 34\%
geometrical coverage. On the outer side of the stainless sphere 200
PMTs of the outer muon veto detector are mounted, these PMTs detects
the Cherenkov light caused by muons passing through.

The sphere contains one thousand tonnes of pseudocumene. Finally,
the innermost core of the facility contains roughly 280 tons of the
scintillating liquid bounded within a 100 $\mu$m thick nylon transparent
bag with $\sim$4.2 m radius. The water and the pseudocumene buffers,
as well as the scintillator itself, have a record-low level of radioactive
purity. The energy of each event is measured using light response
of the scintillator, and the position of the interaction is determined
using timing information from the PMTs. The latter is important for
the selection of the innermost cleanest part of the detector within
3 meters radius, as only the internal 100 tonnes of scintillator have
the radioactive background low enough to allow the solar neutrino
detection, the scintillator layer close to the nylon serves as an
active shield against the $\gamma$ originating from the nylon trace
radioactive contamination. The threshold of the detector is set as
low as possible to exclude triggering from the random dark count of
PMTs. The Borexino has excellent energy resolution for its size, this
is the result of the high light yield of $\sim$500 p.e./MeV/2000
PMTs. The energy resolution is as low as 5\% at 1 MeV.

\section{Data processing}

The low-energy range, namely 165-590 keV, of the Borexino experimental
spectrum has been recently carefully analyzed with a purpose of the
pp-neutrino flux measurement ~\cite{PP14}. The data were acquired
from January 2012 to May 2013 and correspond to 408 live days of the
data taking. These are the data were collected at the beginning of
the second phase of Borexino which had started after the additional
purification of the liquid scintillator following the calibration
campaign of 2010-2011~\cite{Calib12}. The main backgrounds for the
solar neutrino studies were significantly reduced in the Phase 2,
the content of $^{85}$Kr is compatible with zero, and background
from $^{210}$Bi reduced by a factor 3 to 4 compared to the values
observed at the end of the Phase 1 just before the purification. 

The experimental spectrum is presented in Fig.1 The main features
of the experimental spectrum can be seen in the figure: the main contribution
comes from the $^{14}$C decays at low energies (below 200 keV), the
monoenergetic peak corresponds to 5.3 MeV $\alpha-$particles from
$^{210}$Po decay. The statistics in the first bins used in the analysis
is very high, of the order of $5\times10^{5}$events, demanding development
of the very precise model for the studies - the allowed systematic
precision at low energy part should be comparable to the statistical
fluctuations of 0.14\%.

\begin{figure}[!th]
\begin{centering}
\includegraphics[scale=0.6]{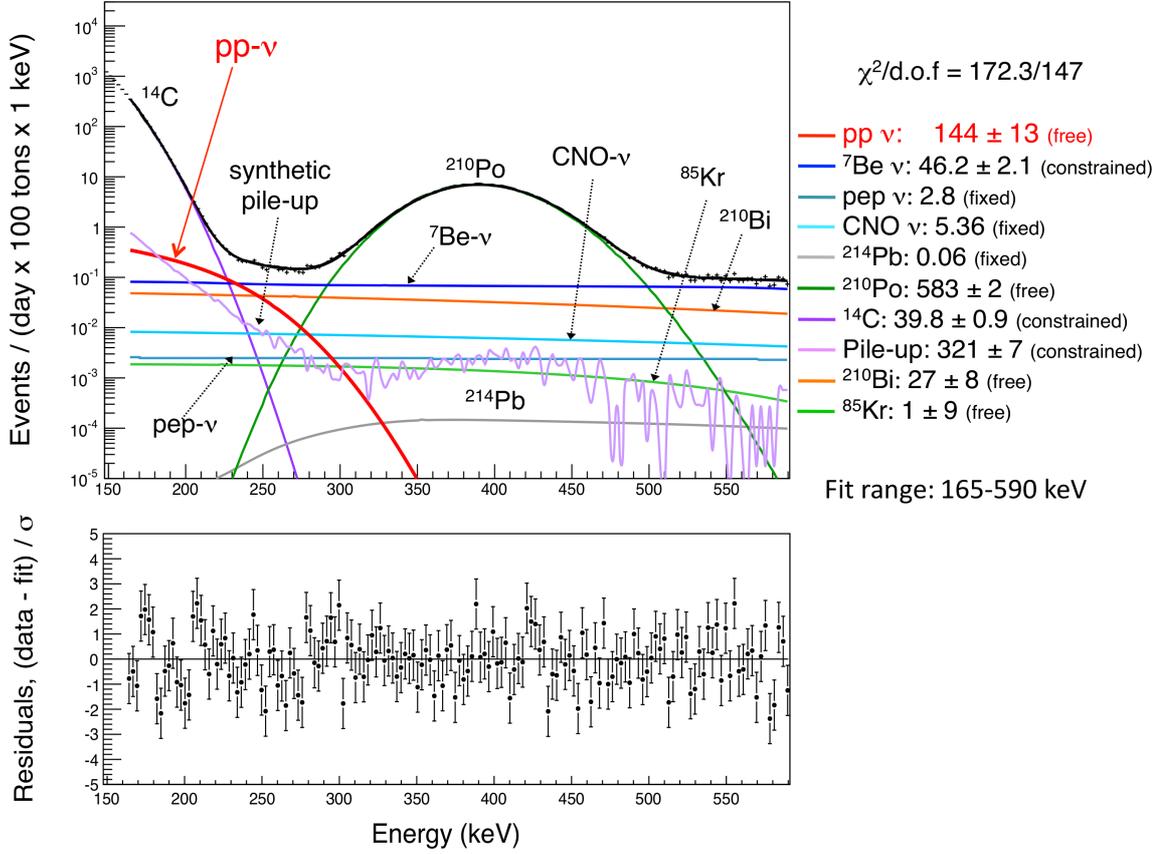}
\par\end{centering}

\caption{Borexino energy spectrum between 165 and 590 keV. The pp neutrino
component is shown in red, the $^{14}$C spectrum in dark purple and
the synthetic pile-up in light purple. The large green peak is $^{210}$Po
$\alpha$-decays. $^{7}$Be (dark blue), pep and CNO (light blue)
solar neutrinos, and $^{210}$Bi (orange) spectra are almost flat
in this energy region. }
\end{figure}

\subsection{Data analysis}

The Borexino spectrum in the low energy range is composed mainly of
the events from $\beta$- decays of \ensuremath{^{14}}C
 present in liquid organic oscillator in trace quantities, its measured
abundance with respect to the $^{12}$C is $(2.7\pm0.1)\times10^{-18}$
g/g. The \ensuremath{\beta}
-decay of \ensuremath{^{14}}
C is an allowed ground-state to ground-state (\ensuremath{0^{+}\rightarrow1^{+}}
) Gamow-Teller transition with an endpoint energy of \ensuremath{E_{0}=156.476\pm0.004}
\ensuremath{\;}
keV. 

In Borexino the amount of the active PMTs is high ($\sim$2000), demanding
setting of the high acquisition threshold in order to exclude detector
triggering from random coincidence of dark count in PMTs: hardware
trigger was set at the level of 25 PMTs in coincidence within 30 ns
window, providing negligible random events count. The acquisition
efficiency corresponding to 25 triggered PMTs is roughly 50\% and
corresponds to the energy release of $\sim$50 keV. In present analysis,
the same as in the ``pp''-analysis, the threshold was set at the
lowest possible value at $\sim$60 triggered PMTs ($\sim$160 keV).
In independent measurement with laser the trigger inefficiency was
found to be below $10^{-5}$ for energies above 120 keV~\cite{Brx14}.

\subsection{Energy resolution}

The most sensitive part of the analysis is the behaviour of the energy
resolution with energy. The variance of the signal is smeared by the
dark noise of the detector (composed of the dark noise from individual
PMTs). In order to account for the dark noise the data were sampled
every two seconds forcing randomly fired triggers. Some additional
smearing of the signal is expected because of the continuously decreasing
number of the PMTs in operation. The amount of live PMTs is followed
in real time and we know precisely its distribution, so in principle
this additional smearing can be precisely accounted for. It was found
that the following approximation works well in the energy region of
interest:

\begin{center}
$\sigma_{N}^{2}=N(p_{0}-p_{1}v_{1})+N^{2}(v_{T}(N)+v_{f})$,
\par\end{center}

where $N=N_{0}<f(t)>_{T}$ is average number of working PMTs during
the period of the data taking, $f(t)$ is a function describing the
amount of working PMTs in time with $f(0)=N_{0}$. The last parameter
here is $v_{f}(N)=<f^{2}(t)>_{T}-<f(t)>_{T}^{2}$, it is the variance
of the $f(t)$ function over the time period of the data taking.

An additional contribution to the variance of the signal was identified,
it is the intrinsic width of the scintillation response. From the
simple consideration this contribution reflects the additional variations
due to the fluctuations of the delta- electrons production and the
energy scale non-linearity. It should scale inversely proportional
to the energy loss. Because of the limited range of the sensitivity
to this contribution, basically restricted to the very tail of the
$^{14}$C spectrum, the precise energy dependence could be neglected
and we used a constant additional term in the resolution. Taking it
all together, the variance of the energy resolution (in terms of the
used energy estimator) is: 

\begin{center}
$\sigma_{N}^{2}=N(p_{0}-p_{1}v_{1})+N^{2}(v_{T}(N)+v_{f})+\sigma_{d}^{2}+\sigma_{int}^{2}$,
\par\end{center}

where $\sigma_{d}$ is contribution of the dark noise (fixed) and
$\sigma_{int}$ is contribution from the intrinsic line shape smearing.
The probability $p_{1}$is linked to the energy estimator with relation
$n=Np_{1}$.

\subsection{Scintillation line shape}

The shape of the scintillation line (i.e. the response of the detector
to the monoenergetic source uniformly distributed over the detector's
volume) is another sensitive component of the analysis. A common approximation
with a normal distribution is failing to describe the tails of the
MC-generated monoenergetic response already at the statistics of the
order of $10^{3}$ events. This was notified already in the first
phase of Borexino and the approximation of the scintillation line
shape with generalized gamma- function~\cite{SM07} have been used
to fit monoenergetic $^{210}$Po peak in the solar $^{7}$Be neutrino
analysis~\cite{Be707,Be708,Be11}. The generalized gamma- function
(GGC) was developed for the energy estimator based on the total collected
charge, but it provided a reasonable approximation for the energy
estimator based on the number of triggered PMTs given the moderate
statistics corresponding to the total amount of the events in $^{210}$Po
peak. The fit quality of the $^{210}$Po peak is rather insensitive
to the residual deviations in the tails. This is not the case for
the precise $^{14}$C spectrum modeling, as all the events in the
fraction of the $^{14}$C spectrum above the energy threshold originate
from the spectral smearing. An ideal detector's response to the point-like
monoenergetic source at the center is a perfect binomial distribution
and it would be well approximated by a Poisson distribution. When
dealing with real response one should adjust the ``base distribution''
width to take into account at least the additional smearing of the
signal due to the various factors. The problem with binomial ``base
function'' (or with its Poisson approximation) is that its width
is defined by the mean value. In case of Poisson the variance of the
signal coincide with mean $\mu$. A better approximation of the response
function was tested with MC model, namely the scaled Poisson (SP)
distribution:

\begin{equation}
f(x)=\frac{\mu^{xs}}{(xs)!}e^{-\mu},\label{ScaledPoisson}
\end{equation}
featuring two parameters, that could be evaluated using expected mean
and variance of the response:

\begin{equation}
s=\frac{\sigma_{n}^{2}}{n}\text{ and }\mu=\frac{n^{2}}{\sigma_{n}^{2}}.\label{Pars}
\end{equation}

The agreement of the approximation and the detector response function
was tested with Borexino MC model and it was found that at low energies
(\ref{ScaledPoisson}) better reproduces the scintillation line shape
compared to the generalized gamma function up to the statistics of
$10^{8}$, while at energies just above the $^{14}$C tail both distribution
gives comparable approximation. The quality of the fit was estimated
using $\chi^{2}$ criterion, for example with $10^{7}$ total statistics
(these events are uniformly distributed in the detector and then the
FV is selected) for $n=50$ (approximately 140 keV) we found $\chi^{2}/n.d.f.$=88.0/61
for the GGC compared to $\chi^{2}/n.d.f.$=59.3/61 for the SP distribution. 

As proven by MC tests, the SP distribution as a base function works
well in the energy region of interest despite of the additional smearing
due to the factors enlisted in the previous paragraph. This is a result
of the ``absorption'' of the relatively narrow non-statistical distributions
by the much wider base function, as follows from MC such an absorption
results in the smearing of the total distribution without changing
its shape. 

As it was noted above, the fit was performed in $n$ scale. All the
theoretical spectra involved in the fit were first translated into
the $n$ scale and then smeared using resolution function (\ref{ScaledPoisson})
with $\mu$ and scale factor $s$ calculated using (\ref{Pars}).
As it is clear from the discussion above, the detector's response
has the shape described by (\ref{ScaledPoisson}) only in the ``natural''
$n$ scale. If the measured values of $n$ would been converted into
the energy, the shape will be deformed because of the non-linearity
of the energy estimator scale with respect to the energy, complicating
the construction of the precise energy response.

\subsection{Standard fit }

The ``standard'' options of the spectral fit are: number of triggered
PMTs in a fixed time window of 230 ns (npmts) used as energy estimator;
62--220 npmts fit range; 75.5$\pm$1.5 tonnes fiducial volume (defined
by the condition R$<$3.02 m and $|Z|<$1.67 m).

The rate of the solar neutrinos is constrained either at the value
found by Borexino in the different energy range ($\mathbf{R({}^{7}Be)=}$
$48\pm2.3$~cpd~\cite{Be11}), or fixed at the prediction of the
SSM in the SMW/LMA oscillation scenario\textbf{ (}R(pep)=2.80 cpd\textbf{,
}R(CNO)=5.36 cpd). All counts here and below are quoted for 100 tonnes
of LS.

The $^{14}$C rate was constrained at the value found in independent
measurement with the second cluster $\mathbf{R(^{14}C})=40\pm1$ Bq
(or $R(^{14}\text{C})=(3.456\pm0.0864)\times10^{6}$~cpd). The synthetic
pile-up rate was constrained at the values found with the algorithm.
The normalization factors for other background components were mainly
left free ($\mathbf{^{85}Kr}$,$\mathbf{^{210}Bi}$ and $\mathbf{^{210}Po}$)
and the fixed rate of $\mathbf{^{214}Pb}$ ($R(\mathbf{^{214}Pb)=0.06}$
cpd) was calculated on the base of the amount of identified radon
events. The light yield and two energy resolution parameters ($v_{T}$
and $\sigma_{int}$) are left free. The position of the $\mathbf{^{210}Po}$
is also left free in the analysis, decoupling it from the energy scale.

\subsection{Systematics study}

An evident source of systematics is uncertainty of the FV. The FV
mass is defined using position reconstruction code, residual bias
in the reconstructed position is possible. The systematic error of
the position reconstruction code was defined during the calibration
campaign, comparing the reconstructed source position with the nominal
one~\cite{PP14,Calib12}. At the energies of interest the
systematic error on the FV mass is 2\%.

The stability and robustness of the measured pp neutrino interaction
rate was verified by performing fits varying initial conditions, including
fit energy range, method of pile-up construction, and energy estimator.
The distribution of the central values for pp-neutrino interaction
rates obtained for all these fit conditions was then used as an estimate
of the maximal systematic error (partial correlations between different
factors are not excluded).

The remaining external background in the fiducial volume at energies
relevant for the pp neutrino study is negligible. In the particular
case of the very low-energy part of the spectrum, the fit was repeated
in five smaller fiducial volumes (with smaller radial and/or z-cut),
which yields very similar results, indicating the absence of the influence
of the external backgrounds at low energies.

\section{Results and Implications}

\begin{figure}[!th]

\begin{centering}
\includegraphics[width=0.8\textwidth]{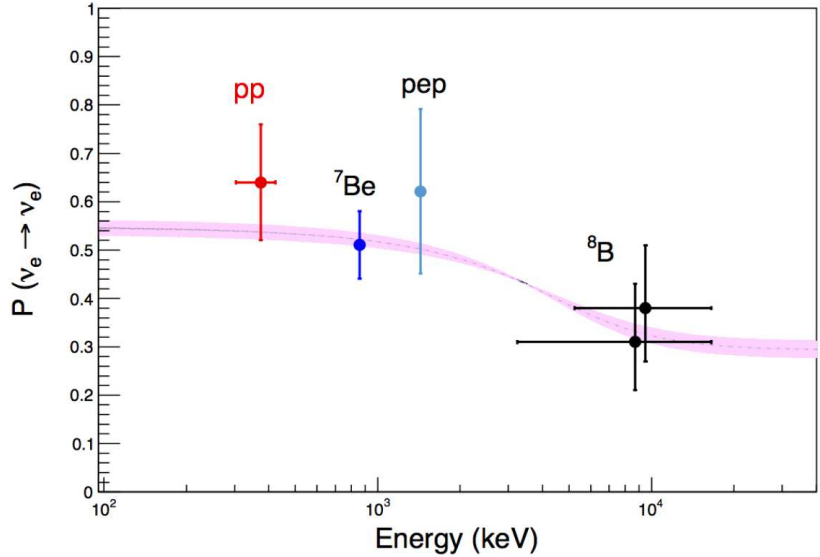}
\par\end{centering}

\caption{\label{Pee-fig}Survival probabilities for electron neutrino (Borexino
only data from~\cite{PP14,Be11,BrxB8,BrxPEP})}

\end{figure}

The solar pp neutrino interaction rate measured by Borexino is $pp=144\pm13(stat)\pm10(syst)$
cpd/100 t, compatible with the expected rate of $pp_{theor}=131\pm2$
cpd/100 t. The corresponding total solar pp-neutrino flux is $\phi_{pp}(Borex)=(6.6\pm0.7)\times10^{10}$
cm$^{-2}$s$^{-1}$, in a good agreement with the combined best fit
value of the radiochemical and other solar experiments $\phi_{pp}(other)=(6.0\pm0.8)\times10^{10}$
cm$^{-2}$s$^{-1}$~\cite{SAGE09}. Both are in agreement with the
expected value of $6.0\times(1.000\pm0.006)\times10^{10}$ cm$^{-2}$s$^{-1}$.

The survival probability for electron neutrino from pp-reaction is
$P_{ee}(Borex)=0.64\pm0.12$. This is the fourth energy range explored
by Borexino, all the Borexino results on the electron neutrino survival
probability are presented graphically in Fig.\ref{Pee-fig}.

Taking into account that Borexino and other experiments measurements
are independent, the results can be combined:

\[
\phi_{pp}=(6.37\pm0.46)\times10^{10}cm^{-2}s^{-1}.
\]

The electron neutrino survival probability measured in all solar but Borexino 
experiment is $P_{ee}(other)=0.56\pm0.08$, combining it with Borexino one we obtain:

\[
P_{ee}=0.60\pm0.07,
\]
 well compatible with theoretical prediction of the MSW/LMA model
$0.561_{-0.042}^{+0.030}$.

\begin{table}
\scriptsize{
\begin{tabular}{|c|c|c|c|c|c|c|}
\hline 
Reaction & GS98~\cite{Grevesse} & AGS09~\cite{Asplund} & Units & Measurement & MeV/1$\nu$ & L\tabularnewline
 &  &  & cm$^{-2}$s$^{-1}$ &  & &  ${\times}10^{26}$ W$\cdot$s$^{-1}$\tabularnewline
\hline 
\hline 
 &  &  &  & $6.0\pm0.8$~\cite{SAGE09} &  & \tabularnewline
pp & $5.98\pm0.04$ & $6.03\pm0.04$ & $\times10^{10}$ & $6.6\pm0.7$ \cite{Brx14} & 13.10 & \tabularnewline
 &  &  &  & $6.37\pm0.46$ &  & $3.76\pm0.28$\tabularnewline
\hline 
pep & $1.44\pm0.012$ & $1.47\pm0.012$ & $\times10^{8}$ & 1.6$\pm0.3$ \cite{BrxPEP} & 11.92 & $0.009\pm0.002$\tabularnewline
\hline 
$^{7}$Be & $5.00\pm0.07$ & $4.56\pm0.07$ & $\times10^{9}$ & $4.87\pm0.24$ \cite{Be11} & 12.60 & $0.276\pm0.014$\tabularnewline
\hline 
$^{8}$B & $5.58\pm0.14$ & $4.59\pm0.14$ & $\times10^{6}$ & $5.25\pm0.16$ \cite{SNO-LETA} & 6.63  & $1.57\pm0.05$\tabularnewline
 &  &  &  & & & $\times{10^{-4}}$\tabularnewline
\hline 
hep & $8.0\pm2.4$ & $8.3\pm2.5$ & $\times10^{3}$ & $<23$\cite{SNO-hep} &  & \tabularnewline
\hline 
$^{13}$N & $2.96\pm0.14$ & $2.17\pm0.14$ & $\times10^{8}$ & CNO: &  & \tabularnewline
\cline{1-4} 
$^{15}$O & $2.23\pm0.15$ & $1.56\pm0.15$ & $\times10^{8}$ & $<7.4$\cite{BrxPEP} &  & \tabularnewline
\cline{1-4} 
$^{17}$F & $5.52\pm0.17$ & $3.40\pm0.16$ & $\times10^{6}$ &  &  & \tabularnewline
\hline 
\end{tabular}
}
\caption{%
\label{SSMvsData}{The Standard Solar Model
predictions (for high metallicity and low metallicity abundances) and current experimental status of the Solar neutrino
fluxes measurement. The limits are given for 90\% C.L.. The corresponding energy release
is calculated in the last column}%
}
\end{table}

All available measurements of the solar neutrino fluxes are shown in Tab.\ref{SSMvsData}. The total energy production in the solar reactions observed till now (by detecting
corresponding neutrino fluxes) is $4.04\pm0.28$ W$\cdot$s$^{-1}$ in a good agreement
with a total measured $L_{\odot}=3.846\times10^{26}$ W$\cdot$s$^{-1}$. There is
not much space left for the unknown energy sources, the 90\% C.L.
lower limit for the total energy production (conservatively assuming zero contribution from the
not-observed reactions) is $L_{tot}=3.68\times10^{26}$ W$\cdot$s$^{-1}$. If one assumes that such an unknown source exists, its total
power with 90\% probability can't exceed $0.15\times10^{26}$ W$\cdot$s$^{-1}$. In other words no more than
4\% of the total energy production in the Sun is left for the
unknown energy sources, confirming that the Sun shines due to the thermonuclear 
fusion reactions.

\section{Acknowledgments}

The Borexino program is made possible by funding from INFN (Italy),
NSF (USA), BMBF, DFG, and MPG (Germany), RFBR: Grants 14-22-03031
and 13-02-12140, RFBR-ASPERA-13-02-92440 (Russia), and NCN Poland
(UMO-2012/06/M/ST2/00426). We acknowledge the generous support and
hospitality of the Laboratori Nazionali del Gran Sasso (LNGS).

\bibliographystyle{unsrt}
\bibliography{pp}

\end{document}